\documentclass{ws-procs9x6-cpt25}

\usepackage{setspace}
\usepackage{wasysym}
\usepackage{epsfig}
\usepackage{epstopdf} 
\usepackage{pstricks}
\usepackage{multirow}
\usepackage{booktabs}
\usepackage{array}
\usepackage{mwe}
\usepackage{soul}
\usepackage{caption}

\begin{document}

\def\thpr{these proceedings}

\def\al{\alpha}
\def\be{\beta}
\def\ga{\gamma}
\def\de{\delta}
\def\ep{\epsilon}
\def\ve{\varepsilon}
\def\ze{\zeta}
\def\et{\eta}
\def\th{\theta}
\def\vt{\vartheta}
\def\io{\iota}
\def\ka{\kappa}
\def\la{\lambda}
\def\vpi{\varpi}
\def\rh{\rho}
\def\vr{\varrho}
\def\si{\sigma}
\def\vs{\varsigma}
\def\ta{\tau}
\def\up{\upsilon}
\def\ph{\phi}
\def\vp{\varphi}
\def\ch{\chi}
\def\ps{\psi}
\def\om{\omega}
\def\Ga{\Gamma}
\def\De{\Delta}
\def\Th{\Theta}
\def\La{\Lambda}
\def\Si{\Sigma}
\def\Up{\Upsilon}
\def\Ph{\Phi}
\def\Ps{\Psi}
\def\Om{\Omega}
\def\mn{{\mu\nu}}

\def\cL{{\cal L}}
\def\lrvec#1{ \stackrel{\leftrightarrow}{#1} }

\def\fr#1#2{{{#1} \over {#2}}}
\def\half{{\textstyle{1\over 2}}}
\def\quar{{\textstyle{1\over 4}}}
\def\eigh{{\textstyle{1\over 8}}}
\def\frac#1#2{{\textstyle{{#1}\over {#2}}}}

\def\prt{\partial}

\def\etal{{\it et al.}}

\def\pt#1{\phantom{#1}}
\def\ol#1{\overline{#1}}

\def\sb{\overline{s}{}}

\def\stx{\sb^{\bar t \bar x}}
\def\sty{\sb^{\bar t \bar y}}
\def\stz{\sb^{\bar t \bar z}}

\newcommand{\beq}{\begin{equation}}
\newcommand{\eeq}{\end{equation}}
\newcommand{\bea}{\begin{eqnarray}}
\newcommand{\eea}{\end{eqnarray}}
\newcommand{\bit}{\begin{itemize}}
\newcommand{\eit}{\end{itemize}}
\newcommand{\rf}[1]{(\ref{#1})}

\def\bt{{\tilde b}}
\def\ct{{\tilde c}}
\def\dt{{\tilde d}}
\def\gt{{\tilde g}}
\def\Ht{{\tilde H}}

\newcommand{\refeq}[1]{(\ref{#1})}
\def\etal {{\it et al.}}

\def\etal {{\it et al.}}

\title{Achieving Full Coverage of the SME Minimal Matter Sector}

\author{Facundo Martin Lopez, Zhiyu Zhang, Bianca Rose Lott, Jay D. Tasson}

\address{Physics and Astronomy Department, Carleton College,\\
Northfield, Minnesota 55057, United States}

\begin{abstract}
Existing experiments and data sets can be leveraged to obtain 
additional sensitivities to Lorentz violation, beyond those originally sought, through a more precise consideration of the boost of the experiment through the background.
In fact,
access to the full coefficient space of the flat-spacetime single-fermion limit of the minimal matter sector of the Standard-Model Extension can be obtained.
In this work we present this coverage for a sample particle in the context of a simplified model of Earth's motion.
\end{abstract}

\bodymatter

\section{Introduction}
Lorentz violation originating from the Planck scale\cite{ks}
has been intensely sought over the past quarter century\cite{datatables}
in the context of an effective field-theory based framework
known as the Standard-Model Extension (SME).\cite{ks1998,kgrav}
Searches in the minimal fermion sector with readily available ordinary matter
particles (protons, neutrons, and electrons)
have been among the most prevalent and sensitive.
Despite this effort,
the 2025 edition of the {\it Data Tables for Lorentz and CPT Violation,}\cite{datatables}
shows a number of Lorentz-violating degrees of freedom yet to be explored in experimental and observational searches.
Our recent work addresses this gap by exploiting higher-precision frame transformations.\cite{jof}
In this work,
we summarize key aspects of our approach
and provide an example of the sensitivities that are possible
in the context of a simplified model of the motion of an Earth-based laboratory.
We emphasize that, while the results presented here in the context
of our simplified model are useful as an illustration and point of comparison with the general results,
those interested in true bounds or the analysis of future experiments
should refer to Ref.~\refcite{jof}.

\section{The fermion sector}

The structure of the fermion sector is discussed in many more comprehensive resources\cite{ks1998,lane,datatables}
and is summarized elsewhere in these proceedings.\cite{jayproc}
Here, we briefly define the relevant variables.
The coefficients for Lorentz violation appearing in the SME Lagrange density
that we explore in this study are
$b_\mu$, $d_\mn$, $g_{\la\mn}$, and $H_\mn$.
These coefficients result in 35 independent Lorentz-violating degrees of freedom per fermion in the minimal flat-spacetime single-fermion limit that we consider here.  They are a subset of the 44 degrees of freedom per fermion typically characterized by the tilde coefficients.
This subset contains the remaining unexplored degrees of freedom associated with protons, neutrons, and electrons.
The explicit form of the tilde coefficients can be found in Table P56 of Ref.~\refcite{datatables}.

In what follows, we consider published results\cite{berlin}
based on the most sensitive class of SME experiments with fermions -- those seeking the effects of the hamiltonian\cite{lane}
\beq
H = \bt_j \si_j,
\label{bdotsi}
\eeq
where $\si_j$ are the Pauli spin matrices
and the tilde coefficient $\bt_j$ is the combination
\beq
\bt_j = b_j -\half\ve_{jkl}H_{kl}-m(d_{jt}-\half\ve_{jkl}g_{klt}).
\label{Eq:btilde}
\eeq
Here $m$ is the mass of the fermion.

\section{Sensitivities from simplified laboratory motions}

\begin{table}[tbph]
\centering\begin{tabular}{|c |c ||c |c|}
\hline
  \multicolumn{1}{|c|}{Coefficient} &
  \multicolumn{1}{c|}{Current} &
  \multicolumn{1}{c||}{Feasible} &
  \multicolumn{1}{c|}{Frequency} \\ \hline\hline

$\bt_{X} $& $10^{-33}$  & - & - \\

$\bt_{Y} $& $10^{-33}$ & - & -  \\

$\bt_{Z} $& 
   $ 10^{-29}$ &
  - & - \\
  
$\bt_{T} $& $10^{-26}$  & $\bf 10^{-29}$ & $\om_L+\Om$ \\

 $\bt^*_{X}$&
 $\boxed{\bf 10^{-12}}$ & $\boxed{\bf 10^{-16}}$ & $\om_L -2\Om$  \\
 
  $\bt^*_{Y} $&
  $\boxed{\bf 10^{-15}}$ & 
   $\boxed{\bf 10^{-16}}$ & $\om_L - 2\Om$ \\
   
  $\bt^*_{Z} $&
  $\boxed{\bf 10^{-15}}$ &
   - & -\\

$\dt_{+} $& $10^{-27}$  & $\bf 10^{-29}$ & $\Om$ \\

$\dt_- $& $10^{-26}$  & $\bf 10^{-29}$ & $\om_L + \Om$ \\

$\dt_{Q} $& $10^{-26}$  & $\bf 10^{-29}$ & $\om_L - \Om$ \\

$\dt_{XY} $&
 $10^{-27}$  & $\bf 10^{-29}$ & $\om_L + \Om$ \\
 
$\dt_{YZ} $&
 $10^{-26}$  & $\bf 10^{-28}$ & $\om_L + \Om$ \\
 
$\dt_{ZX}  $&
 $\boxed{\bf{10^{-27}}}$  & $\boxed{\bf 10^{-29}}$ & $\Om$ \\
 
$\dt_{X} $& $10^{-28}$  & - & - \\

$\dt_{Y} $& $10^{-28}$  & - & - \\

$\dt_{Z} $& $\boxed{\bf 10^{-24}}$  & - & - \\
   
$\Ht_{XT} $&
 $\bf 10^{-27}$  & $\bf 10^{-29}$ & $\Om$ \\
 
$\Ht_{YT} $&
 $\bf 10^{-27}$  & $\bf 10^{-29}$ & $\Om$ \\
 
$\Ht_{ZT} $&
 $10^{-27}$  & $\bf 10^{-29}$ & $\om_L-\Om$
 \\ \hline
\end{tabular}
\caption{Maximum Reach sensitivities to the $\bt_J, \bt^*_J, \dt_\pm, \dt_Q, \dt_{JK}, \dt_J, \Ht_{JT}$ combinations of SME coefficients\cite{datatables} for the neutron,
as demonstrated using simplified laboratory motions.
The ``Current" column lists the new limits that would be obtained from the existing results of searches at the sidereal frequency in the context of a circular-orbit model along with existing limits.\cite{datatables}  Boxed results are sensitivities to coefficients that have never been constrained to date.  Boldface indicates a better sensitivity than the current bounds. We emphasize that these are to be used for illustrative purposes only and do not constitute real bounds, which are instead placed in our more comprehensive work.\cite{jof}   The ``Feasible" column lists estimated sensitivities that could be achieved in a suitable analysis conducted at the listed frequency.  Numeric values are in GeV units.}
\label{AllButg}
\end{table}

In the flat spacetime limit, the SME coefficients can be thought of as fields that are constant
and uniform throughout large regions of the universe.
As such,
they generate time-dependent effects in the lab due to 
the motion of the lab through this background.
A reference frame attached to the Sun is sufficiently inertial over the time scales of relevant experiments. Thus, the coefficients are taken as constants in a standard Sun-centered frame.\cite{datatables}
We use capital indices for components of tensors in this frame.

Sensitive experiments seeking the periodic effects
associated with motions of the Earth-based laboratories
are key tools in searching for Lorentz violation.
In this work, we consider the most significant of these motions:
the boost of the earth as it orbits the Sun ($\be_\oplus\sim10^{-4}$) at the annual frequency $\Om$, which we take as a fixed circular orbit;
the rotation of the Earth at the sidereal frequency $\om$,
which we take as occurring around a fixed axis;
and the boost of the laboratory as it revolves around the Earth ($\be_L\lesssim10^{-6}$) at the sidereal frequency,
which we explore in the context of a perfect-sphere Earth approximation.
We treat the relevant frame transformations,
which are described in detail elsewhere in these proceedings,\cite{jayproc}
perturbatively in the small quantities $\be_\oplus$ and $\be_L$.
We proceed to sufficiently high order in these quantities to obtain sensitivities to all 35 remaining degrees of freedom in this limit of the SME, whereas few prior SME searches have considered boost effects beyond linear order in such a perturbation series.

We first consider a Maximum Reach Analysis\cite{fgt}, which places constraints
on Lorentz violation by assuming one nonzero Lorentz-violating degree of freedom at a time.
The sensitivities that result from a Maximum Reach Analysis 
that uses this model of laboratory motion and published bounds from neutron experiments\cite{berlin}
are found in Tables \ref{AllButg} and \ref{gtable}.
A Coefficient Separation Approach\cite{fgt} considers a family of coefficients together.
Consideration of sufficiently high order in our perturbative expansion offers exciting new opportunities for Coefficient Separation searches as well.\cite{jof,muon1}

\begin{table}[tbph]
\centering\begin{tabular}{|c |c ||c |c|}
\hline
  \multicolumn{1}{|c|}{Coefficient} &
  \multicolumn{1}{c|}{Current} &
  \multicolumn{1}{c||}{Feasible} &
  \multicolumn{1}{c|}{Frequency} \\ \hline\hline

$\gt_{T} $&
 $10^{-27}$  & {$\bf 10^{-29}$} & $\om_L+\Om$ \\
 
$\gt_{c} $&
 ${10^{-27}}$  & {$\bf 10^{-29}$} & $\om_L+\Om$ \\
 
$\gt_{Q} $& $\boxed{\bf 10^{-24}}$  & - & - \\

$\gt_- $& $\boxed{\bf 10^{-24}}$  & - & - \\
 
$\gt_{TX} $& $\boxed{\bf 10^{-24}}$  & - & - \\

$\gt_{TY} $& $\boxed{\bf 10^{-24}}$  & - & - \\

$\gt_{TZ} $& $\boxed{\bf 10^{-24}}$  & - & - \\

$\gt_{XY} $&
 $\boxed{\bf 10^{-19}}$  & $\boxed{\bf 10^{-20}}$ & $\Om$ \\
 
$\gt_{YX} $&
 $\boxed{\bf 10^{-19}}$  & $\boxed{\bf 10^{-20}}$ & $\Om$ \\
 
$\gt_{ZX} $&
 $\boxed{\bf 10^{-18}}$  & $\boxed{\bf 10^{-20}}$ & $\Om$ \\
 
$\gt_{XZ} $&
  $\boxed{\bf 10^{-18}}$  & $\boxed{\bf 10^{-20}}$ & $\Om$ \\
  
$\gt_{YZ} $&
 $\boxed{\bf 10^{-18}}$  & $\boxed{\bf 10^{-20}}$ & $\om_L-3\Om$ \\
  
$\gt_{ZY} $&
 $\boxed{\bf 10^{-18}}$  & $\boxed{\bf 10^{-20}}$ & $\Om$ \\
 
$\gt_{DX} $& $10^{-28}$  & - & - \\

$\gt_{DY}$& $10^{-28}$  & - & - \\

$\gt_{DZ}$& $\boxed{\bf 10^{-24}}$  & - & - 
 \\ \hline
\end{tabular}
\caption{Maximum Reach sensitivities to the $\gt_T, \gt_c, \gt_Q, \gt_-, \gt_{T,J}, \gt_{JK}, \gt_{DJ}$ combinations of SME coefficients\cite{datatables} for the neutron,
as demonstrated using simplified laboratory motions
using the same conventions, conditions, and notation as Table \ref{AllButg}.}
\label{gtable}
\end{table}

\section{Discussion}

The results of our analysis of higher-order boost effects
in the simplified laboratory-motion approximation generate
a useful point of comparison with the results of our more comprehensive treatment.\cite{jof}
We observe that within the simplified model, 
like the more comprehensive treatment,
full-coverage of the remaining degrees of freedom
in the minimal fermion sector is possible
in a Maximum Reach Analysis using the published results
of experiments sensitive to the interaction in Eq.~\refeq{bdotsi}.
Our more comprehensive treatment results in relatively minor
refinements on the basic model used here.\cite{jof}

The necessary calculations to obtain the results in Tables \ref{AllButg} and \ref{gtable}
are quite involved.
To facilitate a greater intuition for how these results come about,
some examples obtained in the context of special limits
of the full coefficient space are presented elsewhere in these proceedings.\cite{jayproc}
In our comprehensive results,\cite{jof}
we consider the electron and proton sector
in addition to the neutron sector.
We also consider the muon sector in other work.\cite{muon1}
The results obtained here in the context of the simplified model of laboratory motion
demonstrate the potential of considering higher-order effects from the boost of the earth
in extending the sensitivities of experimental searches for Lorentz violation.


\end{document}